\documentclass[a4paper,UKenglish,cleveref]{lipics/lipics-v2021}

\usepackage{needspace}
\usepackage{booktabs}

\usepackage{algorithm}
\usepackage{algpseudocode}
\floatname{algorithm}{Algorithm}

\usepackage{tikz}
\usetikzlibrary{arrows.meta}
\usetikzlibrary{positioning}

\newcommand{\ceil}[1]{\left\lceil #1\right\rceil}
\newcommand{\floor}[1]{\left\lfloor #1\right\rfloor}

\newcommand{\GT}{G^{[T]}}

\newcommand{\Coded}{\mathsf{CODE}}
\newcommand{\Uncoded}{\mathsf{ROUTE}}

\DeclareMathOperator{\mincut}{mincut}

\title{On the Completion Time of Broadcast}
\titlerunning{On the Completion Time of Broadcast}

\author{Grigorii Emdin}{EPFL, Lausanne, Switzerland}{}{}{}
\author{Rachid Guerraoui}{EPFL, Lausanne, Switzerland}{}{}{}
\author{Anton Paramonov}{ETH Z\"urich, Zurich, Switzerland}{}{}{}
\authorrunning{G. Emdin, R. Guerraoui, and A. Paramonov}
\Copyright{Grigorii Emdin, Rachid Guerraoui, and Anton Paramonov}



\hideLIPIcs
\nolinenumbers
\makeatletter
\let\@oddfoot\@empty
\makeatother

\AtBeginDocument{%
  \hypersetup{
    pdftitle={On the Completion Time of Broadcast},
    pdfauthor={Grigorii Emdin, Rachid Guerraoui, Anton Paramonov},
    pdfsubject={Optimal coded broadcast and a logarithmic forwarding approximation}
  }%
}

\begin{document}


\maketitle

\begin{abstract}
We study the completion time of broadcasting $k$ packets from a source to every vertex of an undirected $n$-vertex network in which each edge carries one packet in each direction per round. We work in the centralized setting and consider deterministic routing and network coding schemes.

We introduce a quantity $\tau$, computable in polynomial time, that characterizes the broadcast completion time. First, every broadcast scheme, whether routing or coding, requires at least $\tau$ rounds. Second, we give a polynomial-time computable coding scheme that completes the broadcast in exactly $\tau$ rounds, so $\tau$ is precisely the optimal completion time. Third, we give a polynomial-time computable routing scheme that completes the broadcast in $O(\tau \log n)$ rounds. Consequently, the coding gap for broadcast completion time is $O(\log n)$; we also show that it is at least $3/2$. Finally, we show that distributing the packets among several sources at most doubles the completion time, for both routing and coding, so the $O(\log n)$ bound on the coding gap carries over to multi-source broadcast.
\end{abstract}


\section{Introduction}\label{sec:intro} 

Ever since the first message was sent over the ARPANET in 1969, researchers have been working to make the transmission of information through networks more efficient. In 2000, Ahlswede, Cai, Li, and Yeung observed that nodes need not simply forward the packets they relay. Instead, they can transmit suitable \emph{functions} of the data they have received~\cite{ahlswede2000network}. This idea laid the foundation for the field of \emph{network coding}, a rich area that has since deepened our understanding of network flows~\cite{Jaggi05}, homomorphic cryptography~\cite{krohn2004fly}, graph entropy~\cite{riis2006information}, lower bounds for multiplication circuits~\cite{afshani2019lower} and, of course, coding theory~\cite{koetter2008coding}.

Until recently~\cite{HWZ20}, however, network coding was studied almost exclusively through the lens of \emph{throughput}, and the question of \emph{completion time} was largely left aside. Before defining these notions formally, we illustrate the difference between them with a toy example.

\paragraph*{Throughput versus completion time.}
Consider $n$ nodes arranged in a ring, and let $u$ and $v$ be two adjacent nodes. Suppose $u$ wants to send $k$ packets to $v$, and at most one packet can traverse each link per round. There are two edge-disjoint paths between $u$ and $v$: the direct link and the path going around the rest of the ring. Asymptotically, $u$ can therefore deliver two packets per round to~$v$. This rate is the \emph{throughput}. It would be wrong, however, to conclude that $k$ packets can be delivered in $k/2$ rounds. The long path has $n-1$ edges, so no packet sent along it arrives before round $n-1$. When $k \le n-1$, the direct link alone delivers all $k$ packets within $k$ rounds, and the long path is of no use at all. The number of rounds actually needed to deliver all packets is the \emph{completion time}, and it is the quantity we study in this work.

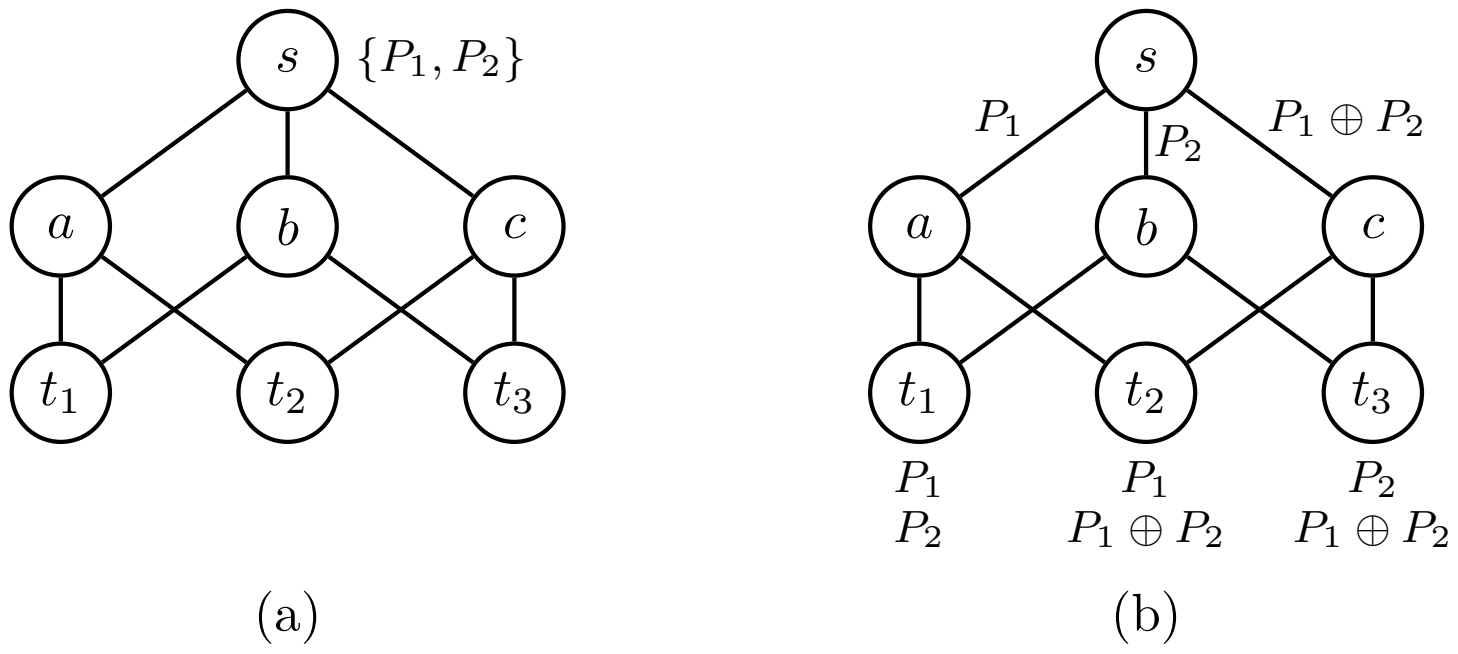
\begin{figure}[ht]
  \centering
  \tikzset{
    vertex/.style = {circle, draw, thick, minimum size=6.5mm, inner sep=0pt},
    link/.style   = {thick},
    lbl/.style    = {font=\footnotesize, inner sep=1pt},
  }%
  \begin{tikzpicture}[x=1.5cm, y=1.1cm]
    \node[vertex] (s)  at ( 0,  0) {$s$};
    \node[lbl, right=2pt of s] {$\{P_1, P_2\}$};
    \node[vertex] (a)  at (-1, -1) {$a$};
    \node[vertex] (b)  at ( 0, -1) {$b$};
    \node[vertex] (c)  at ( 1, -1) {$c$};
    \node[vertex] (t1) at (-1, -2) {$t_1$};
    \node[vertex] (t2) at ( 0, -2) {$t_2$};
    \node[vertex] (t3) at ( 1, -2) {$t_3$};

    \foreach \v in {a, b, c} \draw[link] (s) -- (\v);
    \draw[link] (a) -- (t1);
    \draw[link] (a) -- (t2);
    \draw[link] (b) -- (t1);
    \draw[link] (b) -- (t3);
    \draw[link] (c) -- (t2);
    \draw[link] (c) -- (t3);

    \node at (0, -3.35) {(a)};
  \end{tikzpicture}%
  \hspace{2cm}%
  \begin{tikzpicture}[x=1.5cm, y=1.1cm]
    \node[vertex] (s)  at ( 0,  0) {$s$};
    \node[vertex] (a)  at (-1, -1) {$a$};
    \node[vertex] (b)  at ( 0, -1) {$b$};
    \node[vertex] (c)  at ( 1, -1) {$c$};
    \node[vertex] (t1) at (-1, -2) {$t_1$};
    \node[vertex] (t2) at ( 0, -2) {$t_2$};
    \node[vertex] (t3) at ( 1, -2) {$t_3$};

    \draw[link] (s) -- node[lbl, above left]  {$P_1$}            (a);
    \draw[link] (s) -- node[lbl, right]       {$P_2$}            (b);
    \draw[link] (s) -- node[lbl, above right] {$P_1 \oplus P_2$} (c);

    \draw[link] (a) -- (t1);
    \draw[link] (a) -- (t2);
    \draw[link] (b) -- (t1);
    \draw[link] (b) -- (t3);
    \draw[link] (c) -- (t2);
    \draw[link] (c) -- (t3);

    \node[lbl, below=2pt of t1, align=center] {$P_1$\\ $P_2$};
    \node[lbl, below=2pt of t2, align=center] {$P_1$\\ $P_1 \oplus P_2$};
    \node[lbl, below=2pt of t3, align=center] {$P_2$\\ $P_1 \oplus P_2$};

    \node at (0, -3.35) {(b)};
  \end{tikzpicture}
  \caption{(a) The network. (b) A coding scheme completing dissemination in two rounds: each sink receives two of $P_1$, $P_2$, $P_1 \oplus P_2$ and recovers both packets.}
  \label{fig:coding-example}
\end{figure}

\paragraph*{The power of coding.}
In a ring, nodes can do little more than \emph{routing}, i.e., simply forwarding what they receive. In richer topologies, intermediate nodes can do considerably better. Consider the network in Figure~\ref{fig:coding-example} (a). The source $s$ holds two packets $P_1$ and $P_2$, each $8$ bits long, and every link can carry $8$ bits per round. How many rounds are needed for each of $t_1$, $t_2$, and $t_3$ to learn both $P_1$ and $P_2$?

The reader can verify that if nodes are only allowed to forward data, the task cannot be completed in fewer than $3$ rounds. Suppose instead that, in the first round, $s$ sends $P_1$ to $a$, $P_2$ to $b$, and $P_1 \oplus P_2$ to $c$ (Figure \ref{fig:coding-example} (b)). In the second round, $a$, $b$, and $c$ forward what they received to their neighbors in the third layer. Each sink then holds two of the three values $P_1$, $P_2$, $P_1 \oplus P_2$, from which it can recover both $P_1$ and $P_2$. The dissemination is therefore complete after just two rounds.

Transmitting functions of the available data, rather than merely relaying it, is the essence of network coding. The example shows that for some dissemination problems, network coding achieves strictly lower completion time than routing. The goal of this work is to understand how large this gap can be:
\providecolor{lipicsYellow}{rgb}{0.99,0.78,0.07}%
\begin{center}
  \begin{tikzpicture}
    \node[fill=black!4, rounded corners=3pt, inner xsep=16pt, inner ysep=10pt,
          text width=0.86\linewidth, align=center] (q)
      {\emph{How much can network coding reduce completion time compared to routing?}};
    \fill[lipicsYellow, rounded corners=1.5pt]
      ([xshift=-1.5pt]q.north west) rectangle ([xshift=3pt]q.south west);
  \end{tikzpicture}
\end{center}
We now formally define our setting and then present our results on this question.

\section{Model and problem}\label{sec:model}

An instance $\mathcal{I}$ of the \emph{broadcast problem} consists of a connected undirected graph $G=(V,E)$ with $n=|V|$ vertices, a source $s\in V$, and positive integers $k$ and $B$. We assume that $B \geq \lceil \log_2 n\rceil$. 

\paragraph*{Schemes.}
A solution to the broadcast problem is called a \emph{scheme}. A scheme is a computation initialized with $k$ input packets $P_1,\ldots,P_k\in\{0,1\}^B$ of $B$ bits each, and executed in a deterministic round-based manner as follows.

For every round $t$ and every ordered pair $(v,u)$ with $\{v,u\}\in E$, a scheme specifies the packet that $v$ sends to $u$ in round $t$ as a deterministic function of all packets $v$ received in rounds $1,\ldots,t-1$ and, if $v=s$, of the input packets.
In addition, for every vertex $v$, a scheme specifies a round $T_v\ge0$ and a deterministic output function that maps all packets received by $v$ in rounds $1,\ldots,T_v$ (and, if $v=s$, the input packets) to an output $\mathrm{out}_v$.

We distinguish two kinds of schemes.
\begin{itemize}
  \item In a \emph{coding scheme}, there is no further restriction: a packet may be an arbitrary function of the information available to the sender, as long as it fits into $B$ bits.
  \item In a \emph{routing scheme}, packets are forwarded as opaque tokens. For every round $t$ and every ordered pair $(v,u)$ with $\{v,u\}\in E$, the scheme specifies an index $i\in\{1,\ldots,k\}$. The node $v$ then sends $P_i$ to $u$ in round $t$ if it received $P_i$ in an earlier round (or if $v=s$), otherwise it sends $\bot$.
\end{itemize}

\paragraph*{Completion time and coding gap.}
A scheme is \emph{correct} if, for every input $(P_1,\ldots,P_k)\in(\{0,1\}^B)^k$ and every vertex $v\in V$, the output of $v$ in round $T_v$ is $\mathrm{out}_v=(P_1,\ldots,P_k)$.
The \emph{completion time} of a correct scheme is $T=\max_{v\in V}T_v$.
We denote by $\Uncoded=\Uncoded(\mathcal{I})$ the minimum completion time of a correct routing scheme, and by $\Coded=\Coded(\mathcal{I})$ the minimum completion time of a correct coding scheme.
The \emph{coding gap} of an instance of the broadcast problem is the ratio $\Uncoded/\Coded$, and the broadcast completion time coding gap for $n$-vertex graphs is the supremum of this ratio over all instances with $n$ vertices and all values of $k$ and $B$.

\Needspace{9\baselineskip}
\section{Our results}\label{sec:results}

\textbf{Disclaimer:} Our major results were obtained with substantial assistance from large language models (LLMs). In particular, most of the proofs in this paper were initially generated by OpenAI GPT-6 Pro and were subsequently carefully checked and verified by the authors. The authors take full responsibility for the correctness of all results and arguments presented in the paper.\\

We give a quantity $\tau = \tau(\mathcal{I})$, polynomially computable from a broadcast problem instance $\mathcal{I}$. Informally, $\tau$ is the smallest $T$ such that, in the time-expanded network over $T$ rounds, every receiver is separated from the source only by cuts of capacity at least $k$ (see \cref{sec:overview} for the formal definition). This quantity characterizes broadcast completion time through the following theorems:

\begin{theorem}[Lower Bound]\label{thm:lower}
For every instance $\mathcal{I}$, every correct broadcast scheme takes at least $\tau(\mathcal{I})$ rounds.
\end{theorem}

\begin{theorem}[Routing Upper Bound]\label{thm:forwarding}
For every instance $\mathcal{I}$, there exists a correct routing scheme with completion time $O(\tau(\mathcal{I})\cdot \log n)$. Furthermore, this routing scheme can be computed in polynomial time.
\end{theorem}

Theorems \ref{thm:lower} and \ref{thm:forwarding} together characterize the completion time coding gap of broadcast, thus closing a direction proposed by \cite{HWZ20}.

\begin{theorem}[Broadcast Completion Time Coding Gap]\label{thm:coding-gap}
  The broadcast completion time coding gap is at most $O(\log n)$ and at least $3/2$.
\end{theorem}

The $3/2$ comes from the problem instance as in the toy example in \cref{fig:coding-example}.

Apart from the coding gap, we also characterize the exact round complexity of the optimal coding scheme.

\begin{theorem}[Optimal Coded Broadcast]\label{thm:coded}
For every instance $\mathcal{I}$, there exists a correct broadcast scheme with completion time exactly $\tau(\mathcal{I})$. Furthermore, this scheme only uses linear codes and can be computed in polynomial time.
\end{theorem}
\Cref{thm:lower,thm:coded} together give an exact, polynomially computable characterization of the optimal completion time with coding. This is in sharp contrast to routing: \cite{PW26} showed that determining the exact round complexity of routing is NP-hard. Thus, on the one hand, coding makes a substantial computational difference~--~finding an optimal coding scheme is in $\mathsf{P}$, while finding an optimal routing scheme is NP-hard. On the other hand, by \cref{thm:coding-gap}, the advantage of coding in completion time is at most logarithmic.

Finally, we show that the multi-source broadcast problem, where the $k$ input
packets are distributed among several nodes, is not harder than the single-source problem. The proof of \cref{thm:lower} holds mutatis mutandis, which, coupled with a theorem below, establishes the coding gap for multi-source broadcast to be at most $O(\log n)$.

\begin{theorem}[Multi-Source Broadcast Reduction]\label{thm:multi-source}
  If a broadcast instance $(G, s, k, B)$ possesses a correct routing (resp. coding) scheme with completion time $T$, then a multi-source broadcast instance $(G, \{s_1, \ldots, s_l\}, k, B)$ possesses a routing (resp. coding) scheme with completion time $2T$ for any choice of sources $\{s_1, \ldots, s_l\}$ and any distribution of $k$ packets among them.
\end{theorem}

\paragraph*{Formal verification.}
\Cref{thm:lower,thm:forwarding,thm:coding-gap,thm:coded,thm:multi-source} are formally verified by LLMs in Lean~4 with Mathlib; the formalization is available at \url{https://anonymous.4open.science/r/CodingGapLean-0243/}. The only unproved ingredient is the cited lemma of Chuzhoy, Parter, and Tan~(\cref{lem:cpt}), which enters the routing upper bound as an explicit hypothesis. Polynomial-time computability is not formalized.

\section{Related work}\label{sec:related}

\subsection{Coding gaps for throughput and completion time.}
We provide an overview of state of the art coding gap results for different settings in \cref{tab:gaps}. 

In \emph{unicast}, a source $s$ holds packets that a single receiver $t$ must learn.
In \emph{multicast}, the packets of $s$ must reach every receiver in a set $\mathcal R\subseteq V\setminus\{s\}$; \emph{broadcast} is the special case $\mathcal R=V\setminus\{s\}$.
The \emph{multiple} variants run several such sessions at once: every source in a set $\mathcal S$ holds its own set of packets, destined for its own receiver (multiple unicast) or its own receiver set (multiple multicast/multiple broadcast), and all sessions share the edge capacities.

Each problem admits two modes of consideration.
The \emph{throughput} is the largest rate, in packets per round, at which the unbounded stream of packets can be delivered to all receivers (see \cite{HKL06} for complete definition).
The \emph{completion time} is the number of rounds needed to deliver a fixed set of packets.
For either measure, the coding gap is the ratio between the best routing solution and the best coded solution.

\begin{table}[H]
  \centering
  \small
  \renewcommand{\arraystretch}{1.15}
  \begin{tabular}{@{}lllll@{}}
    \toprule
    Problem & Graph & Gap LB & Gap UB & References \\
    \midrule
    \multicolumn{5}{c}{\textit{Throughput}} \\
    \midrule
    Unicast            & dir./undir. & $1$ & $1$ & \cite{FordFulkerson56} \\
    Broadcast          & dir./undir. & $1$ & $1$ & \cite{Edmonds73,LL04undirected} \\
    Multicast          & undirected  & $8/7$ & $2$ & \cite{LL04undirected,AC04} \\
    Multicast          & directed    & $\Omega(\sqrt{| \mathcal R|})$, $\Omega(n^{0.0418})$ & $|\mathcal R|$ & \cite{AC04,CFS06,LL22} \\
    Multiple multicast & undirected  & $\Omega(\log|\mathcal S|)$, $2^{\tilde{\Omega}(\sqrt{\log \log n})}$ & $O(|\mathcal S|)$, $O(\log n)$ & \cite{BH25} \\
    Multiple unicast   & undirected  & $1$ & $O(\log |\mathcal S|)$ & \cite{LLR95,AR98} \\
    Multiple unicast   & directed    & $\Omega(|\mathcal S|)$ & $O(|\mathcal S|)$ & \cite{LL04unicast} \\
    \midrule
    \multicolumn{5}{c}{\textit{Completion time}} \\
    \midrule
    Unicast            & dir./undir. & $1$ & $1$ & \cite{FF58} \\
    Multiple unicast   & undirected  & $\Omega(\log^c |\mathcal S|)$ & $O(\log^2 |\mathcal S|)$ & \cite{HWZ20} \\
    Broadcast          & undirected  & $3/2$ & $O(\log n)$ & this paper \\
    Multiple broadcast & undirected & $3/2$ & $O(\log n)$ & this paper\\
    \bottomrule
  \end{tabular}
  \caption{Coding gaps. All results assume unit edge capacities. Here $n$ is the number of vertices, $\mathcal S$ is the set of sources, and $\mathcal R$ is the set of receivers of a single-source multicast.}
  \label{tab:gaps}
\end{table}

The work closest to ours is that of Haeupler, Wajc, and Zuzic~\cite{HWZ20}, who study the completion time of multiple unicasts. They formulate a linear program that packs short source--sink paths and pass to its dual, which fractionally cuts these paths: a path is cut if the fractional values of its edges sum to at least one. They then interpret this fractional cut as a \emph{moving cut}, the central concept of their work; see \cref{sec:overview} for how moving cuts compare to the temporal cuts introduced in this paper. This approach does not carry over directly to broadcast. There, the primal would pack shallow trees, and its dual would fractionally cut these trees, a notion with no canonical interpretation.

The most prominent open question in \cref{tab:gaps} is the multiple-unicast conjecture of Li and Li~\cite{LL04unicast} (see also~\cite{HKL06}): "In undirected graphs, the throughput gap for multiple unicast is exactly one."
Besides its intrinsic interest, the conjecture has consequences in complexity theory. Afshani, Freksen, Kamma, and Larsen~\cite{afshani2019lower} showed that it implies that every constant-degree Boolean circuit multiplying two $n$-bit integers has size $\Omega(n\log n)$ -- the lower bound result that is far beyond reach as of now.

\subsection{Broadcast in CONGEST.}

Broadcasting multiple packets has also been studied in the CONGEST model~\cite{Peleg00}, where, unlike in our setting, vertices do not know the network topology in advance.
The classical bound for broadcasting $k$ packets in CONGEST is $\Theta(D+k)$ rounds on graphs of diameter $D$~\cite{Peleg00}.
Ghaffari~\cite{Gha15} observed that this bound is tight only in the worst case: on some graphs of diameter $D$, broadcast can be completed much faster than in $D+k$ rounds.
This raises the challenge of designing a \emph{universally optimal} broadcast algorithm, that is, one whose running time on every graph is close to the best possible on that particular graph.
Exploring this direction, Chandra, Chang, Dory, Ghaffari, and Leitersdorf~\cite{CCDGL24} gave an algorithm tailored to highly connected networks, which broadcasts $k$ packets in $O((n+k)/\lambda)$ rounds, where $\lambda$ is the edge connectivity of the graph.
More recently, Paramonov and Wattenhofer~\cite{PW26} gave an algorithm whose running time is within a polylogarithmic factor of the optimum on every expander graph.

Our results provide the benchmark for this line of research: we now know that "the best possible" for a given graph is precisely $\tau$, rendering a CONGEST algorithm that runs in $\tilde O(\tau)$ rounds universally optimal up to polylogarithmic factors. Although, we do not know yet how to compute $\tau$ in a distributed way.

\section{Techniques overview}\label{sec:overview}
We now proceed to describe, at a high level, the techniques used to obtain our results. Precise proofs are given in \cref{sec:proofs}.
\paragraph*{Temporal cuts.}
A standard way to certify a communication bottleneck is by a cut: the
capacity of an $s$--$v$ cut bounds the amount of flow that can be sent from
$s$ to $v$. For completion time, however, an ordinary cut in the physical
graph is not sufficient, because it does not account for how long the flow
takes to reach the receiver. A path may have ample capacity but be too long
to deliver any information before a given deadline. This suggests applying
the cut argument not to the physical graph itself, but to a graph that also
records time.

This leads naturally to temporal cuts. 
Fix an integer $T\ge0$. The \emph{time-expanded network} $G^{[T]}$
contains a vertex $(v,t)$ for every $v\in V$ and
$t\in\{0,\ldots,T\}$. For every physical edge $\{u,w\}\in E$ and
$t<T$, it contains directed unit-capacity communication arcs
\[
  (u,t)\longrightarrow(w,t+1),
  \qquad
  (w,t)\longrightarrow(u,t+1),
\]
representing the two transmission opportunities on that edge in round
$t+1$. It also contains a directed infinite-capacity storage arc
$(v,t)\to(v,t+1)$, representing the possibility of keeping information
in local memory.

Thus, paths in $G^{[T]}$ represent causal routes through the network,
while cuts represent communication bottlenecks that take the deadline
into account. For $v\neq s$, define
\[
  C_T(s,v)
    =\operatorname{mincut}_{G^{[T]}}\bigl((s,0),(v,T)\bigr),
\]
we call these \emph{temporal cuts}, and let
\[
  \tau
  =
  \min\{T:C_T(s,v)\ge k\text{ for every }v\neq s\}
\]
be the minimal time such that all temporal cuts are at least $k$.

The lower-bound intuition is immediate from the definition of a temporal
cut. If $T<\tau$, then for some receiver $v$ there is a temporal cut of
capacity $c<k$. Only $c$ packet-sized symbols can cross this cut before
the deadline, whereas $v$ has to recover $k$ independent packet-sized
symbols. Hence the cut does not carry enough information to determine
the input. We formalize this counting argument in \cref{sec:lower}.

\textit{Relation to a Moving Cut.} Another way to incorporate time into a cut certificate is through the
\emph{moving cuts} of Haeupler, Wajc, and Zuzic~\cite{HWZ20}. Their
approach assigns artificial lengths to edges, while ours uses an ordinary
cut in the time-expanded graph. Although these viewpoints are conceptually
quite different, perhaps surprisingly, 
the two notions give exactly the same bound; we prove their equivalence in
Appendix~\ref{app:moving-cut}.

We use temporal cuts throughout because we find them more intuitive: they
arise directly from the communication process and make both the lower and
upper bounds transparent. Additionally, unlike moving cuts, the
time-expanded formulation directly allows us to compute an optimal coded
solution.

\paragraph*{Coded upper bound.}
Temporal cuts are also useful for obtaining upper bounds. At $T=\tau$,
every receiver has a temporal min-cut at least $k$. By the max-flow min-cut
theorem, for every $v\neq s$ there is therefore a flow of value $k$ from
$(s,0)$ to $(v,T)$ in the time-expanded network. These flows are obtained
separately for different receivers and need not be simultaneously routable.

This is exactly where network coding becomes useful. The multicast coding
theorem of Jaggi et al.~\cite{Jaggi05} shows that the separate
source--receiver cut conditions are sufficient for multicast: there is a
single linear network code that delivers all $k$ source symbols to every
receiver simultaneously. Applying this theorem to $G^{[\tau]}$ gives a
coded broadcast completing in exactly $\tau$ rounds, as shown in
\cref{sec:coding}.

\paragraph*{Routing upper bound.}
The routing case is more delicate. At $T=\tau$, every receiver $v$ has
temporal min-cut at least $k$, so $k$ units of flow can reach $v$ by time
$T$. However, these flows are obtained separately for different receivers
and may compete for the same edges, so they do not directly give a routing
schedule that serves all receivers simultaneously.

We first extract useful structure for each receiver separately. A classical
theorem of Ford and Fulkerson~\cite{FF58} implies that the maximum amount
that can reach a fixed receiver by time $T$ can be achieved by repeatedly
sending flow along a fixed set of paths in the original graph $G$. In our
unit-capacity setting, these paths can be chosen edge-disjoint. Moreover,
only paths of length at most $T$ can deliver anything by the deadline, and
each such path can deliver at most $T$ packets. Since at least $k$ packets
can reach $v$ by time $T$, there must therefore be at least
$r=\ceil{k/T}$ edge-disjoint $s$--$v$ paths of length at most $T$.

We now have many short paths from $s$ to every vertex, but the path families
for different vertices may overlap heavily. We first use these $s$--$v$ paths 
to obtain strong pairwise connectivity.  Replace every edge of $G$ by two
parallel copies. Fix two vertices $u$ and $v$, and pair their $r$ paths from
$s$. Reverse each $s$--$u$ path and concatenate it with the corresponding
$s$--$v$ path, obtaining $r$ short $u$--$v$ walks. Within the family of $s$--$u$
paths the paths are edge-disjoint, and the
same holds for the family of $s$--$v$ paths. Hence every edge of $G$ is
used at most once by the $s$--$u$ paths and at most once by the
$s$--$v$ paths. Hence every
edge is used by at most two of the resulting walks. The two parallel copies
of each edge therefore suffice to make these walks edge-disjoint; after
erasing cycles, they give $r$ short edge-disjoint $u$--$v$ paths in the
doubled graph. Thus every pair of vertices has many short edge-disjoint
paths in the doubled graph. However, this guarantee is only pairwise: 
the paths obtained for different pairs of vertices may still overlap heavily.

We next use a lemma of Chuzhoy, Parter, and Tan~\cite{CPT20} to turn this
pairwise connectivity into a single collection of short paths with bounded
congestion. Starting with a set $S$ of vertices, the lemma gives a family
of short paths between pairs of vertices in $S$ such that every vertex is
an endpoint of many paths, while every edge of the graph is used by only
constantly many paths.

We use this family to reduce the number of vertices under consideration.
Starting with $S=V$, we can choose a subset $A\subseteq S$ such that
$s\in A$ and $|A|\le\ceil{|S|/2}$. Moreover, for every vertex
$v\in S\setminus A$, we can select $\Theta(r)$ paths from the family given
by the lemma whose other endpoints lie in $A$. Since we only select paths
from the family given by the lemma, the selected paths remain short and
have constant edge congestion. We keep $A$ and repeat the same argument
with $A$ in place of $S$. Since the size of the set decreases by at least
a factor of two at every step, after $O(\log n)$ steps only $s$ remains.

We then use these sets in reverse order for the broadcast. Initially only
$s$ knows all $k$ packets. Suppose that all vertices in a retained set $A$ 
already know all $k$ packets. Every vertex that was removed in the
preceding step has $\Theta(r)$ short paths from vertices of $A$. We divide
its $k$ packets approximately evenly among these paths. Since
$r=\ceil{k/T}$, each path carries only $O(T)$ packets. The paths have
length $O(T)$ and together have constant edge congestion, so the packets
can be pipelined to all new vertices in $O(T)$ rounds. After this phase, 
these vertices also know all $k$ packets.

Thus each level informs a larger set of vertices in $O(T)$ rounds. There
are only $O(\log n)$ levels, so the complete routing schedule takes
$O(T\log n)=O(\tau\log n)$ rounds. See \cref{sec:paths} for the full
construction.

\paragraph*{Multiple sources.}
Our single-source assumption is essentially without loss of generality.
In Appendix~\ref{app:multiple-sources} we show that allowing the $k$
packets to be initially distributed among several sources increases the
completion time by at most a factor of two. The idea is particularly
simple for routing. Suppose we have a $T$-round routing scheme when all
packets start at a single source $s$. For every packet, consider one
causal path by which the original scheme delivers that packet from $s$
to its initial owner in the multiple-source instance. Reversing these
paths in both space and time gathers all $k$ packets at $s$ within
$T$ rounds; we can then run the original broadcast scheme for another
$T$ rounds. Hence the multiple-source instance finishes within $2T$
rounds. For coding schemes one cannot directly reverse the protocol,
since transmitted symbols may be non-invertible functions of several
packets. We handle this in Appendix~\ref{app:multiple-sources} by using
the temporal-cut condition and max-flow min-cut to construct a
$T$-round routing schedule that first gathers all original packets at
$s$, after which the original coded broadcast is executed.

\section{Main Proofs}\label{sec:proofs}

In this section, we first prove \cref{thm:lower} and \cref{thm:coded},
showing that $\tau$ exactly characterizes the optimal completion time for
coded broadcast. We then prove \cref{thm:forwarding}, showing that routing
can complete broadcast in $O(\tau\log n)$ rounds. Combining
\cref{thm:lower,thm:forwarding} yields the coding-gap bound in
\cref{thm:coding-gap}.

\subsection{Lower Bound}\label{sec:lower}

\begin{proof}[Proof of \cref{thm:lower}]
Fix a deadline $T<\tau$.
By the definition of $\tau$, there exists a receiver $v\ne s$ such that $C_T(s,v)<k$.
Since all finite capacities in the time-expanded network are integral, there is a minimum cut from $(s,0)$ to $(v,T)$ of capacity $ c=C_T(s,v)\le k-1$.
Let $U$ denote the source side of this cut, so that $(s,0)\in U$ and $(v,T)\notin U$.

No storage arc can leave $U$, since every storage arc has infinite capacity while the cut has finite capacity $c$. Hence the cut consists of exactly $c$ unit-capacity communication arcs, each corresponding to one communication opportunity in the original network.

For an input $X\in\Sigma^k$, let $Z(X)\in\Sigma^c$
denote the transcript of the symbols transmitted on these $c$ communication arcs, listed in a fixed order.
We claim that the cut transcript $Z(X)$ completely determines the state of every space-time vertex outside $U$, and in particular the state of $(v,T)$, once the protocol and all information fixed before the input $X$ is revealed are given. More precisely, we prove by induction on $t=0,\ldots,T$ that, for every vertex $u$ such that $(u,t)\notin U$, the state
of $u$ at time $t$ is determined by $Z(X)$ and does not require any further
knowledge of the input $X$.

For the base case $t=0$, recall that initially only the source $s$ knows the
input $X$. Since $(s,0)\in U$, every space-time vertex $(u,0)\notin U$ has an
initial state that is independent of $X$. Hence the claim holds at time $0$.

Now suppose that the claim holds at time $t$, and consider any
$(u,t+1)\notin U$. We first observe that $(u,t)\notin U$. Indeed, if
$(u,t)\in U$, then the storage arc $(u,t)\longrightarrow(u,t+1)$ would leave
$U$. Since storage arcs have infinite capacity, this would make the cut
infinite, contradicting the fact that its capacity is $c<\infty$. Thus
$(u,t)\notin U$, and by the induction hypothesis the state of $u$ at time $t$
is determined by $Z(X)$.

It remains to determine the packets received by $u$ during round $t+1$.
Consider an incoming communication arc
$(w,t)\longrightarrow(u,t+1)$.
If $(w,t)\notin U$, then by the induction hypothesis the state of $w$ at time
$t$ is determined by $Z(X)$. Since the protocol is deterministic, the symbol
sent by $w$ on this arc is therefore also determined by $Z(X)$.

If instead $(w,t)\in U$, then the arc
$(w,t)\longrightarrow(u,t+1)$ leaves $U$ and is therefore one of the
communication arcs of the cut. Its transmitted symbol is consequently one
of the entries of the cut transcript $Z(X)$.

Hence the previous state of $u$ and every packet received by $u$ during
round $t+1$ are determined by $Z(X)$. Since the protocol is deterministic,
these quantities uniquely determine the state of $u$ at time $t+1$. This
completes the induction.

Since $(v,T)\notin U$, the induction implies that the state of $v$ at time
$T$, and therefore its output, is completely determined by $Z(X)$. Thus, if
two inputs $X,X'\in\Sigma^k$ satisfy $Z(X)=Z(X')$, then the protocol produces
the same output at $v$ on inputs $X$ and $X'$. Because the protocol must
recover the entire input correctly at $v$, this is possible only if $X=X'$.
Hence the map $X\longmapsto Z(X)$ must be injective.

Let $Q=2^B$ be the number of possible $B$-bit packets. There are $Q^k$
possible inputs in $(\{0,1\}^B)^k$, whereas a cut transcript
$Z(X)\in(\{0,1\}^B)^c$ has at most $Q^c$ possible values. Since
$c\le k-1$, we have $Q^c<Q^k$. Therefore no injective map
from the set of inputs to the set of cut transcripts can exist, a
contradiction.
Thus no correct deterministic protocol can finish by any deadline
$T<\tau$. Consequently, every correct deterministic broadcast protocol
requires at least $\tau$ rounds, and hence $\Coded~\ge~\tau$.
\end{proof}

\subsection{Coding Upper Bound}\label{sec:coding}

\begin{proof}[Proof of \cref{thm:coded}]
We prove \cref{thm:coded} in two steps. First, we show that $\tau$ can be computed in polynomial time. We then set $T=\tau$ and construct a linear network code that completes the broadcast within this deadline.

To compute $\tau$, consider the deadlines $T=1,2,\ldots$ in increasing
order. For each $T$, construct the time-expanded network $\GT$ and replace
every infinite-capacity storage arc by an arc of capacity $k$. This replacement
does not affect whether a source--receiver cut has capacity less than $k$:
any cut containing a storage arc already has capacity at least $k$.

By the max-flow min-cut theorem~\cite{FordFulkerson56}, for every receiver
$v\ne s$, the temporal cut value $C_T(s,v)$ is equal to the value of a maximum
$(s,0)$--$(v,T)$ flow in $\GT$. We therefore compute such a maximum flow for
each receiver and test whether its value is at least $k$. The first deadline
$T$ for which this holds for every $v\ne s$ is exactly $\tau$.

Maximum flow is computable in polynomial time, for example by the
Edmonds--Karp algorithm~\cite{EdmondsKarp72}. By the trivial upper bound
$\tau\le k+n$, only polynomially many deadlines need to be tested.
Since each corresponding time-expanded network has size polynomial in
$n$ and $k$, $\tau$ can be computed deterministically in polynomial time.

We now set $T=\tau$ and construct a linear code attaining this deadline.
We use the following result of Jaggi et al.~\cite[Theorem~3]{Jaggi05}.

\begin{theorem}[Theorem 3 from~\cite{Jaggi05}]
Let $N$ be a unit-capacity directed acyclic multigraph with source $s'$ and
set of sinks $\mathcal{T}$. Let $h$ denote the minimum cut between the source
$s'$ and any sink $t\in\mathcal{T}$. Then there is a deterministic
polynomial-time algorithm that constructs a linear network code multicasting
$h$ source symbols to every sink in $\mathcal{T}$ over a sufficiently large
finite field.
\end{theorem}

We apply this theorem to the time-expanded network at $T=\tau$.
First, replace every infinite-capacity storage arc by an arc of capacity $k$.
This does not change any source--receiver min-cut below $k$.
Next, add a new source $s'$ with $k$ unit-capacity input arcs into $(s,0)$,
and replace every capacity-$k$ storage arc by $k$ parallel unit-capacity arcs.
The resulting network is a unit-capacity directed acyclic multigraph, so we
can apply the theorem above.

Let $\mathcal{T}=\{(v,T):v\ne s\}$ be the set of sinks.
By the definition of $\tau$, the minimum cut from $(s,0)$ to $(v,T)$ has
capacity at least $k$ for every $v\ne s$. Therefore, after adding the new
source $s'$, the minimum cut from $s'$ to every sink $(v,T)$ has capacity at
least $k$. On the other hand, only the $k$ artificial input arcs leave $s'$,
so this minimum cut has capacity at most $k$. Hence the minimum cut from
$s'$ to every sink $(v,T)$ has capacity exactly $k$
Thus the parameter $h$ in the theorem is $k$, and the theorem gives a linear
network code that delivers all $k$ source symbols to every receiver.
For our instance there are $n-1$ receivers, and the deterministic
construction of Jaggi et al. can be carried out over a field of size at least
$n$. Hence, for $Q\ge n$, the required code can be constructed
deterministically in polynomial time.

Finally, install the resulting local encoding coefficients and decoding
matrices before transmission begins. The $k$ artificial input arcs represent
the $k$ source symbols $X_1,\ldots,X_k$ initially available at $s$ and require
no communication rounds.

We execute the code according to the time layers of the time-expanded
network. A communication arc $(u,t)\to(w,t+1)$ carrying a coded symbol
corresponds to node $u$ transmitting that symbol to $w$ during round $t+1$.
The symbol transmitted on such an arc is a linear combination of symbols
available at $(u,t)$, and hence is computable by $u$ before round $t+1$.
A storage arc $(u,t)\to(u,t+1)$ corresponds simply to retaining a symbol in
local memory. Since every arc goes from time layer $t$ to time layer $t+1$,
the code respects causality and can be executed layer by layer without any
additional communication rounds.

At time $T$, the coded information available to every receiver has rank $k$,
so its preinstalled decoding map recovers $X_1,\ldots,X_k$. 
Thus broadcast completes in $T=\tau$ rounds.
\end{proof}

\subsection{Routing Upper Bound}\label{sec:paths}

We now turn to \cref{thm:forwarding}. We first show that the condition
$C_T(s,v)\ge k$ implies the existence of many short, edge-disjoint
paths from $s$ to $v$ in $G$. For this, we use a theorem of
Ford and Fulkerson~\cite{FF58}.

Let $G=(V,E)$ be the original undirected graph of the broadcast
instance $\mathcal I$. Fix a receiver $v\ne s$ and an integer
deadline $T\ge0$. For the moment, consider sending packets only
to $v$, rather than broadcasting them to all vertices.

By the max-flow min-cut theorem~\cite{FordFulkerson56},
$C_T(s,v)$ equals the maximum flow from $(s,0)$ to $(v,T)$
in the time-expanded graph. This is precisely the maximum
number of distinct packets that can be routed from $s$ to $v$
within $T$ rounds in the original graph $G$: communication arcs
represent transmissions, while storage arcs represent waiting
at intermediate vertices.

However, this does not yet give a broadcast scheme, since the
routing solutions for different receivers may use the same edges
at the same times. The first step to overcome this difficulty is
to obtain a more structured collection of paths to each receiver.
For this, we use a theorem of Ford and Fulkerson~\cite{FF58},
which shows that the maximum number of packets deliverable to a
fixed receiver can be achieved by repeatedly sending packets along
a fixed collection of paths. In our setting, these paths can moreover
be chosen edge-disjoint, as we show below. Although paths to
different receivers may still overlap, this structure will allow
us to construct an efficient broadcast scheme.

Let $\bar G=(V,\bar E)$ be obtained from $G$ by replacing each edge
$\{u,w\}$ with the directed arcs $(u,w)$ and $(w,u)$, each of
capacity one. A \emph{flow} assigns an amount $x_e$ to each arc.
It is a feasible $s$--$v$ flow if $0\le x_e\le1$ for every
$e\in\bar E$, the total flow entering each vertex other than $s,v$
equals the total flow leaving it, and its value
$\operatorname{val}(x)=\sum_{(s,u)\in \bar E}x_{su}
-\sum_{(u,s)\in \bar E}x_{us}$
is nonnegative. Let $\mathcal F(\bar G)$ denote the set of feasible
$s$--$v$ flows, and define
\begin{equation}\label{eq:ff-objective}
  J_T(x):=(T+1)\operatorname{val}(x)-\sum_{e\in\bar E}x_e.
\end{equation}

\begin{theorem}[Ford--Fulkerson,
  {\cite[Theorems~3 and~5]{FF58}}]\label{thm:ff}
For every $v\ne s$ and integer $T\ge0$,
\begin{equation}\label{eq:ff-cut-value}
  C_T(s,v)=\max_{x\in\mathcal F(\bar G)}J_T(x).
\end{equation}
Moreover, the maximum is attained by an integral flow.\footnote{
Ford and Fulkerson's Theorem~3 maximizes
$t\operatorname{val}(x)-\sum_{i,j}t_{ij}x_{ij}$,
where $t_{ij}$ is the traversal time of arc $(i,j)$.
Their Theorem~5 shows that, for $t=T+1$, this maximum equals
the amount of flow deliverable within $T$ time periods.
Since every arc in our network has traversal time one,
and flows over time correspond to flows in the time-expanded
network, this gives \eqref{eq:ff-cut-value}.
Their construction yields an integral optimum when the
capacities are integral.}
\end{theorem}

The objective $J_T$ counts the packets delivered by this repeated-path
strategy. To see this, suppose an integral flow $x$ with no cycles is
decomposed into paths $P_1,\ldots,P_a$, each carrying one unit of flow.
Let $\ell_i\le T$ be the length of $P_i$. Since a packet takes
$\ell_i$ rounds to traverse $P_i$, it must depart no later than time
$T-\ell_i$ to arrive by the deadline. Thus, we can send one new
packet along $P_i$ at each time $0,1,\ldots,T-\ell_i$. Since
$\operatorname{val}(x)=a$ and
$\sum_{e\in\bar E}x_e=\sum_{i=1}^{a}\ell_i$, the number delivered is
\[
  \sum_{i=1}^{a}(T+1-\ell_i)
  =(T+1)\operatorname{val}(x)-\sum_{e\in\bar E}x_e
  =J_T(x).
\]

In the next lemma, we show that an integral maximizer of $J_T$ yields
edge-disjoint paths in the original undirected graph $G$, after
removing paths that contribute nothing by the deadline.
Each remaining path delivers at most $T$ packets. Therefore, if
$C_T(s,v)\ge k$, there must be at least $\ceil{k/T}$
such paths.

\begin{lemma}[Short paths from the source]\label{lem:short}
If $T\ge1$ and $C_T(s,v)\ge k$, then $G$ contains at least
$\ceil{k/T}$ edge-disjoint $s$--$v$ paths, each of length at most
$\min\{T,n-1\}$. Such paths can be found in polynomial time.
\end{lemma}

\begin{proof}
Choose an integral flow $x$ in $\bar G$ maximizing $J_T(x)$ in
\eqref{eq:ff-cut-value}, and write $a=\operatorname{val}(x)$.

The arcs carrying positive flow contain no directed cycle. Otherwise,
subtracting one unit around such a cycle would preserve feasibility and
the flow value, while decreasing $\sum_e x_e$ and hence increasing
$J_T(x)$, a contradiction. In particular, the two opposite arcs
corresponding to an edge of $G$ cannot both carry flow, since they
would form a directed cycle of length two.

Decompose $x$ into $a$ paths from $s$ to $v$, each carrying one unit
of flow. No cycle flows remain, by the preceding argument. Denote the
paths by $P_1,\ldots,P_a$ and their lengths by
$\ell_1,\ldots,\ell_a$. Every arc has capacity one, and the two
opposite arcs of an edge of $G$ cannot both be used. Therefore the
paths are edge-disjoint in $G$. They are simple, so $\ell_i\le n-1$ for every $i$.
No $P_i$ has length greater than $T+1$. Indeed, removing its unit of flow
would decrease $\operatorname{val}(x)$ by one and decrease
$\sum_e x_e$ by $\ell_i$, changing $J_T(x)$ by
$\ell_i-(T+1)$. If $\ell_i>T+1$, this would increase $J_T(x)$,
contrary to its maximality. A path of length exactly $T+1$ contributes
zero to $J_T(x)$ and can be discarded. Let $\mathcal P$ be the
remaining family. Every $P\in\mathcal P$ has length at most
$\min\{T,n-1\}$, and
\[
  k
  \le C_T(s,v)
  =J_T(x)
  =\sum_{i=1}^{a}(T+1-\ell_i)
  =\sum_{P\in\mathcal P}(T+1-|P|)
  \le T\,|\mathcal P|.
\]
The last inequality uses $|P|\ge1$, since $s\ne v$.
Thus $|\mathcal P|\ge\ceil{k/T}$.

It remains to show that the required paths can be found in polynomial
time. The integral maximizer $x$ can be computed by a minimum-cost
circulation. Put $m=|E|$. Give every arc of $\bar G$ capacity one
and cost one, and add a distinct artificial return arc
$e_0=(v,s)$ of capacity $m$ and cost $-(T+1)$.
Every feasible flow in $\bar G$ has value at most
$\deg_G(s)\le m$. Thus extending a flow $x$ by setting
$f(e_0)=\operatorname{val}(x)$ gives a feasible circulation.
Conversely, restricting any feasible circulation to the arcs of
$\bar G$ gives a flow of value $f(e_0)$. The circulation cost is
\[
  \sum_{e\in\bar E}f(e)-(T+1)f(e_0)=-J_T(x).
\]
Therefore a minimum-cost circulation corresponds exactly to a maximizer
of $J_T$. A standard polynomial-time minimum-cost circulation algorithm
returns an integral optimum. Decompose this flow as above, discard paths
of length $T+1$, and retain any $\ceil{k/T}$ remaining paths. The
auxiliary network has $n$ vertices and $2m+1$ arcs, and its numerical
data have $O(\log(m+1)+\log(T+1))$ bits. Hence the required paths can be
found in polynomial time.
\end{proof}

We now complete the proof of \cref{thm:forwarding}. Fix
\begin{equation}\label{eq:parameters}
  T=\tau(G,s,k),\qquad
  r=\ceil{k/T},\qquad
  h=\min\{T,n-1\},\qquad
  q=\ceil{r/2}.
\end{equation}
Since $C_0(s,v)=0$ for $v\ne s$, the assumptions $n\ge2$ and
$k\ge1$ imply $T\ge1$. By \cref{lem:short}, for each $v\ne s$ we
can compute a family $\mathcal R_v$ of $r$ edge-disjoint simple paths
from $s$ to $v$, each of length at most $h$. Fix these families for
the rest of the construction. In particular,
$1\le r\le\deg_G(s)\le n-1$, and
\begin{equation}\label{eq:batch}
  k\le rT\le2qT,
  \qquad
  b:=\ceil{k/q}\le2T.
\end{equation}
Paths belonging to different families $\mathcal R_v$ may share edges.
We therefore cannot use all these paths simultaneously without checking
how often each edge is used. Instead, we use them to construct new
path families with a bound on the number of paths using each edge.

An undirected multigraph, which may have parallel edges, is
\emph{$(r,d)$-connected} if every pair of distinct vertices is joined
by $r$ edge-disjoint paths of at most $d$ edges each. We use the
following result.

\begin{lemma}[Chuzhoy--Parter--Tan,
  {\cite[Lemma~6.6]{CPT20}
  \footnote{
Lemma~6.6 states existence. The explicit polynomial-time algorithm follows
from its proof.}
}]\label{lem:cpt}
Let $H$ be a finite undirected $(r,d)$-connected multigraph without
self-loops. Let $U\subsetneq V(H)$ and $s\in V(H)\setminus U$.
There is a family $\mathcal P$ of simple paths, each joining two
different vertices of $U\cup\{s\}$, satisfying
\[
\begin{aligned}
  \bigl|\{P\in\mathcal P:u\text{ is an endpoint of }P\}\bigr|
    &\ge r &&(u\in U),\\
  |P|&\le2d &&(P\in\mathcal P),\\
  \bigl|\{P\in\mathcal P:e\in E(P)\}\bigr|
    &\le2 &&(e\in E(H)).
\end{aligned}
\]
\end{lemma}

Intuitively, the lemma says that if every pair of vertices is connected
by many short edge-disjoint paths, then we can find a single collection
of short paths connecting the selected vertices, such that every vertex
in $U$ is an endpoint of at least $r$ paths, while each edge is used by
at most two paths. These paths need not connect directly to $s$, and
they may pass through other selected vertices. Throughout the proof,
we count paths separately even if they have the same endpoints or
follow the same edges.

Our paths so far connect $s$ to every other vertex, whereas
\cref{lem:cpt} requires short connections between every pair of vertices.
We obtain this property by replacing each edge of $G$ with two parallel
copies, obtaining a graph $2G$. We claim that $2G$ is
$(r,2h)$-connected.
Fix distinct $u,v\ne s$, and pair the paths in $\mathcal R_u$ with
those in $\mathcal R_v$ arbitrarily. For each pair, follow the first
path backwards from $u$ to $s$, then follow the second path from $s$
to $v$. This gives $r$ $u$--$v$ walks of length at most $2h$.
Every edge of $G$ appears at most once in the paths of $\mathcal R_u$
and at most once in the paths of $\mathcal R_v$. Thus it is traversed
at most twice across the resulting walks. Erase cycles from each walk,
without changing its endpoints. For each edge used twice by the remaining
paths, assign its two uses to different copies in $2G$. The paths are
then edge-disjoint in $2G$ and still have length at most $2h$.
For a pair containing $s$, the paths already computed suffice.
This proves that $2G$ is $(r,2h)$-connected.

Now fix any set $S\subseteq V$ with $s\in S$ and $|S|\ge2$.
Apply \cref{lem:cpt} to the doubled graph, taking
$U=S\setminus\{s\}$ and $d=2h$.
For the polynomial-time construction, we use the previously computed
families $\mathcal R_u$, $u\in S\setminus\{s\}$, which remain
edge-disjoint when transferred to $2G$ and have lengths at most $h\le2h$.

In the paths returned by the lemma, replace every edge copy by its
original edge in $G$. Their vertex sequences do not change, so the
paths remain simple and have the same endpoints. If different returned
paths become identical in $G$, continue to count them separately.
We obtain, in polynomial time, a family $\mathcal P_S$ satisfying
\begin{equation}\label{eq:terminal-paths}
\begin{aligned}
  \bigl|\{P\in\mathcal P_S:u\text{ is an endpoint of }P\}\bigr|
    &\ge r &&(u\in S\setminus\{s\}),\\
  |P|&\le4h &&(P\in\mathcal P_S),\\
  \bigl|\{P\in\mathcal P_S:e\in E(P)\}\bigr|
    &\le4 &&(e\in E).
\end{aligned}
\end{equation}
The last bound holds because each edge of $G$ has two copies, and each
copy belongs to at most two returned paths. Every path in
$\mathcal P_S$ joins different vertices of $S$.

We next use these paths to keep a smaller set of vertices containing
$s$, while giving each removed vertex enough paths to the vertices
we keep. The argument follows the cut construction
in~\cite[Corollary~6.2]{CPT20}; we prove the exact
statement needed here.

\begin{lemma}[Reducing the set of vertices]\label{lem:halving}
For every $S\subseteq V$ with $s\in S$ and $|S|\ge2$, the path
families $\mathcal R_v$ allow us to construct, deterministically in
polynomial time, a set $A\subsetneq S$ such that
\[
  s\in A,\qquad |A|\le\ceil{|S|/2},
\]
and, for each $v\in S\setminus A$, exactly $q$ selected simple paths
from vertices of $A$ to $v$. Every selected path has length at most
$4h$, and each edge of $G$ belongs to at most four selected paths in
total.
\end{lemma}

\begin{proof}
Construct $\mathcal P_S$ as above. Form an auxiliary multigraph $F$
on $S$, placing one edge between the endpoints of each path in
$\mathcal P_S$. Different paths with the same endpoints give different
parallel edges. There are no self-loops, since every path joins two
different vertices. In particular, $\deg_F(u)\ge r$ for every
$u\in S\setminus\{s\}$.

Start with any partition $(X,Y)$ of $S$ into two sets. Repeatedly
move a vertex to the other set whenever this strictly increases the
number of edges between $X$ and $Y$, counting parallel edges separately.
For a vertex $u$, let $c_F(u)$ be the number of its incident edges
that cross the current cut. Moving $u$ changes the cut size by
$\deg_F(u)-2c_F(u)$. Therefore, when no move increases the cut size,
\[
  c_F(u)\ge\ceil{\deg_F(u)/2}
  \quad(u\in S).
\]
Every vertex other than $s$ consequently has at least
$\ceil{r/2}=q$ edges crossing the cut.

If $|X|=|Y|$, keep the side containing $s$ and call it $A$.
Otherwise, keep the smaller side and add $s$ if necessary. In the
unequal-size case, the smaller side has at most
$\floor{(|S|-1)/2}$ vertices, so
\[
  |A|\le\floor{(|S|-1)/2}+1=\ceil{|S|/2}.
\]
The same bound holds when the sides have equal size. Since $|S|\ge2$,
we have $|A|<|S|$, and our choice ensures $s\in A$.
If $s$ is moved from the other side into $A$, the number of edges to
$A$ cannot decrease for any vertex remaining outside $A$: edges from
that vertex to $s$ now cross the cut. Thus every $v\in S\setminus A$
still has at least $q$ edges to $A$ in $F$.

For each $v\in S\setminus A$, select any $q$ of these edges and
use the corresponding paths in $G$, directed from $A$ towards $v$.
Each edge crossing the cut has exactly one endpoint outside $A$, so
no path is selected for two different receivers. We have only removed
paths from $\mathcal P_S$. Therefore their lengths remain at most
$4h$, and each edge of $G$ is still used by at most four selected
paths, by~\eqref{eq:terminal-paths}.

The cut can be found in polynomial time. Every path is nonempty, and
each edge belongs to at most four paths, so
\[
  |E(F)|=|\mathcal P_S|
  \le\sum_{P\in\mathcal P_S}|P|
  =\sum_{e\in E}\bigl|\{P\in\mathcal P_S:e\in E(P)\}\bigr|
  \le4|E|.
\]
Every move increases the integer cut size by at least one, and the cut
size is at most $|E(F)|$. Hence there are at most $4|E|$ moves.
At each step, we can find a vertex whose move increases the cut size
by checking all vertices of $F$. Selecting the paths after the cut is
found also takes polynomial time.
\end{proof}

Starting from $S_0=V$, repeatedly apply \cref{lem:halving} to $S_i$
and let $S_{i+1}$ be the set $A$ that we keep. Stop when only $s$
remains. This gives
\begin{equation}\label{eq:hierarchy}
  V=S_0\supsetneq S_1\supsetneq\cdots\supsetneq S_L=\{s\},
  \qquad L\le\ceil{\log_2 n}.
\end{equation}
Indeed, induction gives $|S_i|\le\ceil{n/2^i}$, using
$\ceil{\ceil{x}/2}=\ceil{x/2}$. Every $S_i$ contains $s$, so by
$i=\ceil{\log_2 n}$ it can contain only $s$.
For every step $i<L$, keep the selected paths from $S_{i+1}$ to
$S_i\setminus S_{i+1}$. Each vertex of $S_i\setminus S_{i+1}$
has exactly $q$ such paths, each of length at most $4h$, and each edge
is used by at most four paths in this step. These bounds apply separately
to each step. We remove vertices only from the sets under consideration,
not from $G$: all paths still use the original graph.

\begin{proof}[Completion of the proof of \cref{thm:forwarding}]
Process the steps in reverse order, $L-1,L-2,\ldots,0$.
At the start of step $i$, we require that every vertex of $S_{i+1}$
already has all $k$ packets. This holds for the first step because
$S_L=\{s\}$. We show that step $i$ gives all $k$ packets to every
vertex of $S_i$, establishing the same condition for the next step.

For each receiver $v\in S_i\setminus S_{i+1}$, order its $q$
selected paths and divide the packet indices $\{1,\ldots,k\}$ into
$q$ groups of size at most $b=\ceil{k/q}$. For example, assign
indices congruent to $j$ modulo $q$ to the $j$th path, for
$j=1,\ldots,q$. The groups contain original packets; no packet is
split into smaller pieces. Each selected path starts at a vertex of
$S_{i+1}$, which already has every packet in the group assigned to
that path. Empty groups require no transmissions.

Divide the rounds used for this step into blocks of four consecutive
rounds. For each edge $e$, assign the selected paths using $e$
different round numbers $\sigma_e(P)\in\{1,2,3,4\}$ within a block.
This is possible because at most four selected paths use the edge.
For a selected path $P=(x_0,\ldots,x_\ell)$, list its assigned
packets in increasing index order as $(m_{P,1},\ldots,m_{P,b_P})$.
For $1\le j\le b_P$ and $1\le d\le\ell$, send $m_{P,j}$ from
$x_{d-1}$ to $x_d$ in
\begin{equation}\label{eq:calendar}
  \text{block }j+d-1,
  \qquad
  \text{round }\sigma_{\{x_{d-1},x_d\}}(P)
  \text{ within that block}.
\end{equation}
Thus the transmission takes place in round
$4(j+d-2)+\sigma_{\{x_{d-1},x_d\}}(P)$ of this step, with rounds
numbered from one. Make no other useful transmissions.

We check by induction on $d$ that each sender has the packet when it
is required to send it. For $d=1$, the first vertex of the path has
the packet at the start of the step. For $d>1$, the preceding edge
sends the same packet in block $j+d-2$, so it arrives by time
$4(j+d-2)$, measured from the start of the step. The next transmission
departs at time
$4(j+d-2)+\sigma_{\{x_{d-1},x_d\}}(P)-1$, which is no earlier.
Hence the sender already has the packet. This remains true even when
the path is assigned different round numbers on consecutive edges.

The transmission rule also respects edge capacities. Since $P$ is
simple, each edge appears at most once on $P$. If it is the $d$th
edge, then in block $f$ the only possible transmission along it for
this path has packet index $j=f-d+1$. Thus a path sends at most one
packet on an edge in each block. Different selected paths using that
edge have different assigned rounds. Therefore at most one selected
packet crosses an edge of $G$ in any round, even counting both
directions together. This satisfies the requirement of at most one
packet per direction per round.

A path with a nonempty group finishes by block
$b_P+\ell-1\le b+4h-1$. Allocate
$\Delta=4(b+4h-1)$ rounds to the step, using the model's predetermined
dummy symbol in unused transmission opportunities. Every new receiver
then has all its assigned groups, and hence all $k$ packets.
By~\eqref{eq:batch} and $h\le T$,
\begin{equation}\label{eq:phase}
  \Delta=4(b+4h-1)
  \le4(2T+4T-1)
  \le24T.
\end{equation}
A vertex inside a selected path need not have all $k$ packets at the
start of the step. It only forwards each assigned packet after receiving
it from the preceding vertex. When different receivers need copies of
the same packet, the transmissions along their selected paths are
counted separately, so the edge-capacity check includes all these copies.

After step $i$, every vertex of $S_i$ has all $k$ packets. The next
step can therefore be performed in the same way. By induction, all
vertices have all packets after the $L$ steps. Run the steps one after
another without overlap; step $i$ starts at time $(L-1-i)\Delta$.
It follows from~\eqref{eq:hierarchy} and~\eqref{eq:phase} that
\[
  \Uncoded(G,s,k)
  \le L\Delta
  \le24\tau(G,s,k)\ceil{\log_2 n}.
\]

We finish by checking the deterministic polynomial-time construction.
Find $T=\tau$ by binary search over $\{1,\ldots,k+n-2\}$. For
each tested deadline $t$, compute maximum flows in $G^{[t]}$ to
check whether $C_t(s,v)\ge k$ for every $v\ne s$.
Once these inequalities hold for a deadline, they also hold for every
later deadline, because flow that has arrived can remain at its receiver.
The upper search limit is valid because sending the packets successively
along a spanning tree completes broadcast within $k+n-2$ rounds.
For each test, replace every infinite storage capacity by $k$. This
preserves whether the minimum cut is at least $k$: a cut using a storage
arc already has capacity at least $k$, and every other cut is unchanged.
Each tested network has size polynomial in $n$ and $k$, so the search
takes polynomial time.

Next, use \cref{lem:short} to compute the families $\mathcal R_v$ of
$r$ short paths from $s$ to each receiver. Listing all these paths
requires $O(nrh)$ edge entries, with $r,h\le n-1$, so the doubled
graph and the path lists have polynomial size. Using these paths,
\cref{lem:cpt} and the procedure in \cref{lem:halving} construct each
step in polynomial time. There are at most $\ceil{\log_2 n}$ steps.
\end{proof}

\begin{proof}[Proof of \cref{thm:coding-gap}]
The upper bound $O(\log n)$ follows from
\cref{thm:lower,thm:forwarding}. The lower bound $3/2$ is given by the
instance in \cref{fig:coding-example}.
\end{proof}

\newpage
\bibliography{bibliography}


\newpage
\appendix

\section{Multiple sources}\label{app:multiple-sources}

In this section we show that allowing the input packets to be distributed
among several sources increases the completion time by at most a factor
of two. The statement holds both for routing and for coding.

Formally, let $S=\{s_1,\ldots,s_\ell\}\subseteq V$ be a set of sources.
A distribution of the $k$ packets among these sources is described by a
map
\[
    \sigma:\{1,\ldots,k\}\to S,
\]
where packet $P_j$ is initially available at $\sigma(j)$.
Every vertex must recover the entire tuple $(P_1,\ldots,P_k)$.

We now prove \cref{thm:multi-source}.

\begin{proof}
Fix an arbitrary placement
$\sigma:\{1,\ldots,k\}\to S$.
We first gather all $k$ packets at the original source $s$ within
$T$ rounds. We then run the given single-source broadcast scheme for
another $T$ rounds.
The gathering phase uses only routing, even if the given broadcast
scheme uses coding.

Since the given single-source scheme finishes in $T$ rounds,
\cref{thm:lower} implies that
\[
    C_T(s,v)\ge k
    \qquad\text{for every }v\neq s.
\]
Consider the time-expanded network $G^{[T]}$ and replace every
infinite-capacity storage arc by an arc of capacity $k$.
This does not change any cut of capacity smaller than $k$:
a cut containing a storage arc already has capacity at least $k$,
while every cut containing no storage arc is unchanged.
We also have a min-cut of at least $k$ from $(s,0)$ to $(s,T)$:
the storage arcs
$
    (s,0)\to(s,1)\to\cdots\to(s,T)$
form a path of capacity $k$.
Hence
\begin{equation}\label{eq:multisource-cuts}
    \mincut\bigl((s,0),(v,T)\bigr)\ge k
    \qquad\text{for every }v\in V.
\end{equation}

Add a new sink $z$.
For every packet $P_j$, add a distinct unit-capacity arc
\[
    e_j:(\sigma(j),T)\longrightarrow z.
\]
If several packets have the same initial source, these are simply
parallel arcs.

We claim that every cut separating $(s,0)$ from $z$ has capacity at
least $k$.
Let $U$ be the source side of such a cut.
If all vertices $(\sigma(j),T)$ belong to $U$, then all $k$ arcs
$e_1,\ldots,e_k$ cross the cut, so its capacity is at least $k$.
Otherwise, $(\sigma(j),T)\notin U$ for some $j$.
Then the arcs of the original time-expanded network leaving $U$
already separate $(s,0)$ from $(\sigma(j),T)$.
By~\eqref{eq:multisource-cuts}, they have total capacity at least $k$.

Thus every $(s,0)$--$z$ cut has capacity at least $k$.
On the other hand, the total capacity of the arcs entering $z$ is
exactly $k$.
By the integral max-flow min-cut theorem, there is therefore an
integral $(s,0)$--$z$ flow of value $k$.
All $k$ arcs $e_j$ must be saturated.

Decompose this integral flow into $k$ unit paths.
For every $j$, exactly one of these paths ends with $e_j$.
Deleting that final arc gives a path
\[
    \pi_j:(s,0)\leadsto(\sigma(j),T).
\]
The paths $\pi_1,\ldots,\pi_k$ collectively respect all capacities
of the time-expanded network.
We now reverse these paths in both space and time.
A communication arc
$
    (u,t)\longrightarrow(v,t+1)
$
is replaced by
$    (v,T-t-1)\longrightarrow(u,T-t).
$
Such an arc exists because $G$ is undirected and both directions of
every physical edge have the same unit capacity.
Similarly, a storage arc
$
    (u,t)\longrightarrow(u,t+1)
$
is replaced by the storage arc
$
    (u,T-t-1)\longrightarrow(u,T-t).
$
Thus time reversal is a capacity-preserving bijection on the arcs of
the time-expanded network.

Consequently, the reversed paths are still jointly feasible, and the
reversal of $\pi_j$ is a path from
$    (\sigma(j),0)
    \ \text{to}\
    (s,T).
$
Route the original packet $P_j$ along this path.
Communication arcs correspond to forwarding the packet in the indicated
round, while storage arcs correspond to keeping it in local memory.
Since the reversed paths respect the capacities of the time-expanded
network, no directed physical edge carries more than one packet in any
round.

Therefore all $k$ packets reach $s$ within $T$ rounds.
At that point, $s$ possesses the complete input
$(P_1,\ldots,P_k)$.
We now execute the original single-source broadcast scheme, shifted by
$T$ rounds.
It finishes within another $T$ rounds, so the total completion time is
at most $2T$.

The gathering phase only forwards original packets.
Hence, if the original single-source scheme is a routing scheme, the
entire $2T$-round construction is a routing scheme.
If the original scheme is a coding scheme, the first phase uses routing
and the second phase uses that coding scheme, so the resulting scheme is
a coding scheme.
\end{proof}

\section{Temporal cuts and moving cuts}\label{app:moving-cut}
We compare temporal cuts with the moving cuts of Haeupler, Wajc, and
Zuzic~\cite{HWZ20}. For simplicity, we prove their equivalence for a
single source--receiver pair. The proof extends naturally to sets of
sources and receivers, as in the original definition of moving cuts.

\begin{definition}[Moving cut {\cite[Definition~1.6]{HWZ20}}]
\label{def:moving-cut}
Fix distinct vertices $s,v\in V$. A \emph{moving cut} is an assignment
$\ell:E\to\mathbb Z_{\ge1}$ of positive integer lengths to the edges.
Its \emph{capacity} is
\[
  b(\ell)=\sum_{e\in E}(\ell_e-1).
\]
It has \emph{distance at least $D$} if $d_\ell(s,v)\ge D$, where
$d_\ell$ denotes shortest-path distance with respect to $\ell$.\footnote{%
The original definition permits edge capacities $c_e$ and several
source--sink pairs $(s_i,t_i)$. Its capacity is
$\sum_e c_e(\ell_e-1)$, and its distance condition is
$d_\ell(s_i,t_j)\ge D$ for every $i,j$. Here every $c_e=1$ and there
is only the pair $(s,v)$. The lengths are part of the cut certificate;
they do not change the physical transmission times.}
\end{definition}

\begin{lemma}[Equivalence of temporal and moving cuts]
\label{lem:temporal-moving}
For every $v\ne s$ and integer $T\ge0$,
\begin{equation}\label{eq:temporal-moving}
  C_T(s,v)
  =
  \min_{\substack{
    \ell:E\to\mathbb Z_{\ge1}\\
    d_\ell(s,v)\ge T+1
  }} b(\ell).
\end{equation}
\end{lemma}

\begin{proof}
Write $[x]_+=\max\{x,0\}$. We convert a cut of either type into
one of the other type without increasing its capacity.

\emph{From a temporal cut to a moving cut.}
Let $U$ be the source side of a minimum cut from $(s,0)$ to $(v,T)$
in $G^{[T]}$. Its capacity is finite, so no storage arc leaves $U$.
Consequently, if $(u,t)\in U$, then every later time-copy of $u$ also
belongs to $U$. Thus there are integers $a_u\in\{0,\ldots,T+1\}$
such that
\[
  (u,t)\in U \quad\Longleftrightarrow\quad t\ge a_u.
\]
Here $a_u=T+1$ means that no time-copy of $u$ belongs to $U$.
Since $(s,0)\in U$ and $(v,T)\notin U$, we have
$a_s=0$ and $a_v=T+1$.

For a physical edge $\{u,w\}$, the communication arc
$(u,t)\to(w,t+1)$ leaves $U$ precisely when
$a_u\le t\le a_w-2$. There are $[a_w-a_u-1]_+$ such arcs.
The opposite direction contributes $[a_u-a_w-1]_+$, so
\begin{equation}\label{eq:temporal-label-capacity}
  C_T(s,v)
  =\sum_{\{u,w\}\in E}[|a_u-a_w|-1]_+.
\end{equation}
Set
\[
  \ell_{\{u,w\}}=\max\{1,|a_u-a_w|\}.
\]
Then $b(\ell)=C_T(s,v)$. Moreover, every $s$--$v$ path $P$
satisfies
\[
  \sum_{e\in E(P)}\ell_e
  \ge \sum_{\{u,w\}\in E(P)}|a_u-a_w|
  \ge |a_v-a_s|
  =T+1.
\]
Hence $d_\ell(s,v)\ge T+1$, proving that the right-hand side
of~\eqref{eq:temporal-moving} is at most $C_T(s,v)$.

\emph{From a moving cut to a temporal cut.}
Conversely, let $\ell:E\to\mathbb Z_{\ge1}$ satisfy
$d_\ell(s,v)\ge T+1$. Define
\[
  a_u=\min\{T+1,d_\ell(s,u)\},
  \qquad
  U=\{(u,t):0\le t\le T,\ t\ge a_u\}.
\]
The labels are integers, with $a_s=0$ and $a_v=T+1$.
Therefore $U$ contains $(s,0)$, excludes $(v,T)$, and no storage
arc leaves $U$. By the same counting argument as above, the capacity
of this temporal cut is
\[
  \sum_{\{u,w\}\in E}[|a_u-a_w|-1]_+.
\]
Since $G$ is undirected, shortest-path distances satisfy
$|d_\ell(s,u)-d_\ell(s,w)|\le\ell_{\{u,w\}}$.
Truncating both distances at $T+1$ cannot increase their difference,
so $|a_u-a_w|\le\ell_{\{u,w\}}$. Consequently,
\[
  C_T(s,v)
  \le \sum_{\{u,w\}\in E}[|a_u-a_w|-1]_+
  \le \sum_{e\in E}(\ell_e-1)
  =b(\ell).
\]
Minimizing over all admissible $\ell$ gives the reverse inequality
in~\eqref{eq:temporal-moving} and completes the proof.
\end{proof}

\end{document}